\documentclass[twocolumn,epjc3]{svjour3}
\smartqed  
\usepackage{amssymb,amsmath}
\usepackage{graphicx}
\usepackage{hyperref}

\newcommand\rf[1]{(\ref{eq:#1})}
\newcommand\lab[1]{\label{eq:#1}}
\newcommand\nonu{\nonumber}
\newcommand\br{\begin{eqnarray}}
\newcommand\er{\end{eqnarray}}
\newcommand\be{\begin{equation}}
\newcommand\ee{\end{equation}}

\newcommand\foot[1]{\footnotemark\footnotetext{#1}}
\newcommand\lb{\lbrack}
\newcommand\rb{\rbrack}

\newcommand\bc{\begin{center}}
\newcommand\ec{\end{center}}

\newcommand\partder[2]{\frac{{\partial {#1}}}{{\partial {#2}}}}

\renewcommand\a{\alpha}

\renewcommand\d{\delta}

\newcommand\eps{\epsilon}
\newcommand\vareps{\varepsilon}

\newcommand\h{\frac{1}{2}}
\renewcommand\k{\kappa}
\renewcommand\l{\lambda}
\renewcommand\L{\Lambda}
\newcommand\m{\mu}
\newcommand\n{\nu}

\newcommand\vp{\varphi}
\renewcommand\P{\Phi}
\newcommand\pa{\partial}

\newcommand\pr{\prime}

\renewcommand\th{\theta}

\newcommand\wti{\widetilde}

\newcommand\cB{{\mathcal B}}

\newcommand\cH{{\mathcal H}}

\newcommand{\ct}[1]{\cite{#1}}
\newcommand{\bib}[1]{\bibitem{#1}}
\newcommand\PRL[3]{\textsl{Phys. Rev. Lett.} \textbf{#1}, #3 (#2)}

\newcommand\PRD[3]{\textsl{Phys. Rev.} \textbf{D#1}, #3 (#2)}

\newcommand\PLB[3]{\textsl{Phys. Lett.} \textbf{#1B}, #3 (#2)}
\newcommand\CQG[3]{\textsl{Class. Quantum Grav.} \textbf{#1}, #3 (#2)}

\newcommand\AoP[3]{\textsl{Ann. of Phys.} \textbf{#1}, #3 (#2)}

\newcommand\IJMPA[3]{\textsl{Int. J. Mod. Phys.} \textbf{A#1}, #3 (#2)}

\newcommand\vpdot{\stackrel{.}{\varphi}}
\newcommand\vpddot{\stackrel{..}{\varphi}}

\newcommand\adot{\stackrel{.}{a}}
\newcommand\addot{\stackrel{..}{a}}

\newcommand\cBdot{\stackrel{.}{\mathcal B}}

\journalname{Eur. Phys. J. C}

\begin{document}

\title{Dark Energy and Dark Matter From Hidden Symmetry of Gravity Model with a
Non-Riemannian Volume Form}
\titlerunning{Dark Matter From Hidden Symmetry}

\author{Eduardo Guendelman\thanksref{e1,addr1}
        \and
        Emil Nissimov\thanksref{e2,addr2} 
	\and 
	Svetlana Pacheva\thanksref{e3,addr2} 
}

\thankstext{e1}{e-mail: guendel@bgu.ac.il}
\thankstext{e2}{e-mail: nissimov@inrne.bas.bg}
\thankstext{e3}{e-mail: svetlana@inrne.bas.bg}

\institute{Department of Physics, Ben-Gurion University of the Negev, Beer-Sheva, 
Israel \label{addr1}
          \and
Institute for Nuclear Research and Nuclear Energy,
Bulgarian Academy of Sciences, Sofia, Bulgaria \label{addr2}
}


\date{Received: date / Accepted: date}

\maketitle

\begin{abstract}

We show that dark energy and dark matter can 
be described simultaneously by ordinary 
Einstein gravity interacting with a single scalar field provided the scalar field
Lagrangian couples in a symmetric fashion to two different spacetime
volume-forms (covariant integration measure densities) on the spacetime
manifold -- one standard Riemannian given by $\sqrt{-g}$ (square-root of the
determinant of the pertinent Riemannian metric) and another non-Riemannian
volume-form independent of the Riemannian metric, defined in terms of an
auxiliary antisymmetric tensor gauge field of maximal rank.
Integration of the equations of motion of the latter auxiliary gauge field
produce an {\em a priori} arbitrary integration constant that plays the 
role of a {\em dynamically generated} cosmological constant or dark energy.
Moreover, the above modified scalar field action turns out to possess a
{\em hidden} Noether symmetry whose associated conserved current describes a
pressureless ``dust'' fluid which we can identify with the dark matter
completely decoupled from the dark energy. 
The form of both the dark energy and dark matter that results from above class
of models is insensitive to the specific form of the scalar field
Lagrangian. By adding appropriate perturbation, which breaks the above
hidden symmetry and along with this it couples dark matter and dark energy, 
we also suggest a way to obtain growing dark energy in the
present universe's epoch without evolution pathologies. 



\end{abstract}

\section{Introduction}
\label{intro}


The late time Universe appears to be dominated by two components, both of them 
``non-luminous'' or ``dark''. The  dominant contribution, about 70\% of the 
energy density of the universe is well described by a cosmological constant term, 
as introduced originally by Einstein and has been also given the name ``dark energy''.
This contribution  leads to gravitational repulsion. The cosmological constant 
or dark energy is not diluted by the expansion of the universe. 
The other subdominant contribution, about 25\%  of the energy density of the 
universe is well described by a pressureless fluid, which is called ``dark matter''.
As opposed to the dark energy it is gravitational attractive and it gets diluted 
by the universe expansion, it can form structures, \textsl{etc}.

Dark energy was observationally discovered rather recently through the 
observation of type {\em Ia} supernova \ct{dark-energy-observ}. 

Dark matter was first postulated in the 1930s, separately by J. Oort and F. Zwicky, 
due to the anomaly of the orbital velocity of some stars in the Milky Way galaxy
and the orbital velocity of galaxies in clusters. A recent review of dark
matter is given in Ref.\ct{dark-matter-rev}, reviews of dark energy can be found in
\ct{dark-energy-rev} and a review of both dark matter and dark energy in 
\ct{dark-matter-energy-rev}.

In this paper we study a class of models providing a unified description of
dark energy and dark matter starting from a well-defined gravity-scalar-field 
Lagrangian action constructed by means of both standard Riemannian as
well as an alternative non-Riemannian (\textsl{i.e.}, independent of the
pertinent Riemannian metric) volume forms (covariant integration
measure densities) on the spacetime manifold.
The introduction of such modified ``two-measure'' gravity-matter theory
(the general class of ``two-measure'' gravity  models was originally proposed
in Refs.\ct{TMT-orig})
opens the possibility to obtain both dark energy and dark matter from a
single scalar field, as it was already observed in \ct{eduardo-singleton}. 
This was further generalized in \ct{eduardo-ansoldi} by the inclusion of another 
field with phantom-like kinetic energy so as to produce growing dark energy. 
In the present paper we will achieve growing dark energy in a different way that 
does not invoke phantom kinetic terms and without introducing additional fields.

In a recent paper \ct{cruz-etal} a model providing unifying description of
dark energy and dark matter was proposed by studying thermodynamics of
cosmological systems where a constraint on the pressure being a constant was
introduced from the very beginning. In the present case we start from a
well-defined Lagrangian action principle for a modified gravity-scalar-field 
system which produces systematically the constant pressure constraint
in a self-consistent dynamical way as part of the pertinent equations of
motion.

Here we will proceed to discover the fundamental reasons how modified
gravity-matter models, generalizing those studied in Ref.\ct{eduardo-singleton},
succeed to describe simultaneously both dark matter and dark energy. 
We find that this is realized due to:
\begin{itemize}
\item
(i) The existence of a {\em hidden} 
(strongly nonlinear)
Noether symmetry of the underlying
single scalar field Lagrangian, that implies a conservation 
law from which it follows that there is conserved current giving rise to the
dark matter component.
\item
(ii) An {\em a priori} arbitrary integration constant appears in a dynamical 
constraint on the scalar field Lagrangian, which plays the role of a 
{\em dynamically generated cosmological constant}
and provides the dark energy component. The fact that the latter arises from an 
integration constant makes the observed vacuum energy density totally decoupled 
from the parameters of the matter Lagrangian.
\end{itemize}

Both fundamental features (i)-(ii) arise in a way completely independent of the 
specific form of the scalar field Lagrangian and the details of the scalar field 
dynamics.

Other treatments that unify dark energy and dark matter have appeared before, 
for example, the Chaplygin gas models \ct{chaplygin-kamen,chaplygin-bilic}. 

More recently, a ``mimetic'' dark matter model was proposed \ct{mimetic-grav}
based on a special covariant isolation of the conformal degree of
freedom in Einstein gravity, whose dynamics mimics cold dark matter as a
pressureless ``dust''. Also, the cosmological implications of the
``mimetic'' matter were studied in some detail (second Ref.\ct{mimetic-grav}).
For further generalizations and extensions of ``mimetic'' gravity, see
Refs.\ct{mimetic-grav-extend}.

Models of explicitly coupled dark matter and dark energy described in terms
of two different scalar fields were proposed in Ref.\ct{copeland}.

As a final introductory remark let us briefly describe the usefulness 
of employing the formalism based on alternative non-Riemannian spacetime volume-forms,
\textsl{i.e.}, alternative covariant integration measure densities in gravity-matter 
Lagrangian actions independent of the pertinent Riemannian metric. The latter
have profound impact in any field theory models with general coordinate 
reparametrization invariance -- general relativity and its extensions, strings and
(higher-dimensional) membranes as already studied in a series of previous
papers on this subject \ct{TMT-orig,mstring,TMT-recent}.

Although formally appearing as (almost) ``pure-gauge'' dynamical degrees of freedom 
\foot{For a detailed canonical Hamiltonian analysis a'la Dirac of
gravity-matter theories with several independent non-Riemannian spacetime volume-forms,
see \ct{quintess} and Appendix A in \ct{buggy}; see also Section 2 below for the
simple case of one non-Riemannian volume form.}
the non-Riemannian volume-form fields trigger a number of remarkable physically 
important phenomena:
\begin{itemize}
\item
Non-Riemannian volume-form formalism in gravity-matter theories 
naturally generates a {\em dynamical cosmological constant} as an arbitrary 
dimensionful integration constant.
At this point it resembles the earlier proposed unimodular gravity 
formulated as a fully generally covariant theory within the framework of
Dirac's constraint Hamiltonian method 
\ct{henneaux-teitelboim} \foot{The original idea of unimodular gravity is in 
Einstein's works \ct{einstein-unimodular}; in
more modern context  it appeared in \ct{ng-zee}.}.
Unimodular gravity became further an object of active studies -- for the latest 
developments, especially path integral quantization, equivalence with the
fully diffeomorphism invariant formulation, and further references, see 
\ct{bufalo-oksanen-tureanu}.
On the other hand, the non-Riemannian volume-form approach goes well beyond the
dynamical cosmological constant generation and has significantly broader 
scope. Namely, unimodular gravity in its generally covariant form
(Eq.(18) in \ct{henneaux-teitelboim}, which appears as a particular case of a
gravity theory with a non-Riemannian volume-form) is equivalent to standard
general relativity (on classical level, except that the cosmological
constant is an integration constant). On the other hand, generic
non-Riemannian\--volume\--form\--modified gravity theories are non-trivial 
extensions to general relativity; see also the next points here below.
\item
Employing two different non-Riemannian volume-forms
generates several independent arbitrary integration constants leading
to the construction of a
new class of gravity-matter models, which produce an effective scalar potential with 
{\em two infinitely large flat regions} \ct{emergent,quintess}.
This allows for a unified description of both 
early universe inflation as well as of present dark energy epoch.
\item
A remarkable feature is the existence of a stable initial phase of
{\em non-singular} universe creation preceding the inflationary phase
-- stable ``emergent universe'' without ``Big-Bang'' \ct{emergent}.
\item
Within non-Riemannian-modified-measure minimal $N=1$ supergravity the 
dynamically generated cosmological constant triggers spontaneous supersymmetry
breaking and mass generation for the gravitino (supersymmetric 
Brout-Englert-Higgs effect) \ct{susyssb}. Applying the same non-Riemannian
volume-form formalism to anti-de Sitter supergravity allows to produce
simultaneously a very large physical gravitino mass and a very small 
{\em positive} observable cosmological constant \ct{susyssb} in accordance 
with modern cosmological scenarios for slowly expanding universe of the present epoch
\ct{dark-energy-observ}.
\item
Adding interaction with a special nonlinear (``square-root'' Maxwell) gauge field 
(known to describe charge confinement in flat spacetime) produces various
phases with different strength of confinement and/or with deconfinement, 
as well as gravitational electrovacuum ``bags'' partially mimicking the properties 
of {\em MIT bags} and solitonic constituent quark models (for details, see \ct{buggy}).
\end{itemize}

In Section 2 we briefly describe the basics of the non-Riemannian
volume-form (modified measure) approach, including elucidating the meaning of the
dynamically generated cosmological constant (\textsl{i.e.}, dark energy appearing 
as an arbitrary integration constant in a dynamical constraint on the scalar field
Lagrangian) from the point of view of the canonical Hamiltonian formalism. 
In Section 3 we derive the hidden symmetry and the associated Noether conserved 
current of the present modified-measure gravity-scalar-field model leading
to the ``dust-fluid'' interpretation of a part of the scalar field energy
density, \textsl{i.e.}, dark matter.
In Section 4 few implications for cosmology are considered. We briefly discuss 
perturbing our modified-measure gravity-scalar-field model which breaks the above 
crucial hidden symmetry
and triggers 
(upon appropriate choice of the perturbation) a growing dark energy in the
present day universe' epoch without invoking any pathologies of
``cosmic doomsday'' or future singularities kind \ct{doomsday}.
Our concluding remarks are contained in the last Section 5. 

\section{Gravity-Matter Formalism With a Non-Riemannian Volume-Form}
\label{TMT}

Our starting point is the following non-conventional gravity-scalar-field
action (for simplicity we use units with the Newton constant $G_N = 1/16\pi$):
\be
S = \int d^4 x \sqrt{-g}\, R + \int d^4 x \bigl(\sqrt{-g}+\P(B)\bigr) L(\vp,X) \; ,
\lab{TMT}
\ee
which in fact is a simple particular case of the general class of the so called 
``two-measure'' gravity-matter theories proposed more than a decay ago \ct{TMT-orig}.
The notations we are using are as follows:
\begin{itemize}
\item
The first term in \rf{TMT} is the standard Hilbert-Einstein action;
$\sqrt{-g} \equiv \sqrt{-\det\Vert g_{\m\n}\Vert}$ is the standard
Riemannian integration measure density with $g_{\m\n}$ being the standard 
Riemannian spacetime metric.
\item
$\P(B)$ denotes an alternative non-Riemannian generally covariant
integration measure density defining an alternative non-Riemannian volume
form on the pertinent spacetime manifold:
\be
\P(B) = \frac{1}{3!}\vareps^{\m\n\k\l} \pa_\m B_{\n\k\l} \; ,
\lab{mod-measure}
\ee
where $B_{\m\n\l}$ is an auxiliary maximal rank antisymmetric tensor gauge
independent of the Riemannian metric. 

$B_{\m\n\l}$ \rf{mod-measure} will also be called 
``measure gauge field''\foot{In the original papers \ct{TMT-orig} an alternative
parametrization of $B_{\m\n\l}$ through 4 auxiliary scalar fields 
$\{\phi^I\}_{I=1,\ldots, 4}$ was used --  
$B_{\m\n\l}=\frac{1}{4} \vareps_{IJKL}\,
\phi^I \pa_\m \phi^J \pa_\n \phi^K \pa_\l \phi^L$, so that
$\Phi (B) = \frac{1}{4!} \vareps^{\m\n\k\l} \vareps_{IJKL}\,
\pa_\m \phi^I \pa_\n \phi^J \pa_\k \phi^K \pa_\l \phi^L 
= \det\Vert \frac{\pa \phi^I}{\pa x^\m} \Vert$.
In a recent study \ct{struckmeier} of general
relativity as an extended canonical gauge theory a similar Jacobian
representation of the covariant integration measure has appeared in terms of
additional scalar fields. However, unlike the present case in the
construction of Ref.\ct{struckmeier} the additional scalar fields enter 
also in the proper Lagrangian.}.
\item
$L(\vp,X)$ is general-coordinate invariant Lagrangian of a single scalar field 
$\vp (x)$ of a generic ``k-essence'' form \ct{k-essence}:
\br
L(\vp,X) = \sum_{n=1}^N A_n (\vp) X^n - V(\vp) \; ,
\lab{k-essence-L} \\
X \equiv - \h g^{\m\n}\pa_\m \vp \pa_\n \vp \; ,
\nonu
\er
\textsl{i.e.}, a nonlinear (in general) function of the scalar kinetic term $X$.
\end{itemize}

Varying \rf{TMT} w.r.t. $g^{\m\n}$, $\vp$ and $B_{\m\n\l}$ yield the
following equations of motion, respectively:
\br
R_{\m\n} -\h g_{\m\n} R = \h T_{\m\n} \; ,
\lab{einstein-eqs} \\
T_{\m\n} = g_{\m\n} L(\vp,X) + 
\Bigl( 1+\frac{\P(B)}{\sqrt{-g}}\Bigr) \partder{L}{X} \pa_\m \vp\, \pa_\n \vp \; ;
\lab{EM-tensor}
\er
\be
\partder{L}{\vp} + \bigl(\P(B)+\sqrt{-g}\bigr)^{-1}
\pa_\m \Bigl\lb\bigl(\P(B)+\sqrt{-g}\bigr) 
g^{\m\n}\pa_\n \vp \partder{L}{X}\Bigr\rb = 0 \; ;
\lab{vp-eqs}
\ee
\be
\pa_\m L (\vp,X) = 0 \;\;\; \longrightarrow \;\;\;
L (\vp,X) = - 2M = {\rm const} \; ,
\lab{L-const}
\ee
where $M$ is arbitrary integration 
constant\foot{Dynamical constraints like the one on the scalar field Lagrangian
in Eq.\rf{L-const}, which routinely appear in all instances 
of applying the non-Riemannian volume-form method in gravity-matter theories,
resemble at first sight analogous constraints on scalar field Lagrangians in 
the ``Lagrangian multiplier gravity'' \ct{LMG}. We would like to point out
that this formalism is in fact a special particular case of the more general 
approach based on non-Riemannian spacetime volume-forms, which appeared
around a decade earlier \ct{TMT-orig}. Dynamical constraints in the latter approach result
from the equations of motion of the auxiliary ``measure'' gauge fields and,
thus, they always involve {\em arbitrary integration constants} like 
$M$ in Eq.\rf{L-const}, as opposed to picking some {\em a priori} fixed constant
within the ``Lagrange multiplier gravity'' formalism. For further advantages
of the non-Riemannian volume-form formalism, see the final remarks in the
Introduction.}
(the factor $2$ is for later convenience).

Already at this point it is important to stress that the scalar field dynamics 
is determined entirely by the first-order differential equation - the 
dynamical constraint
Eq.\rf{L-const} ($X - V(\vp) = - 2M$ in the simplest case of \rf{k-essence-L}). 
The standard second order differential equation \rf{vp-eqs} is in fact a consequence
of \rf{L-const} together with the energy-momentum conservation 
$\nabla^\m T_{\m\n} = 0$.

The physical meaning of the ``measure'' gauge field $B_{\m\n\l}$ \rf{mod-measure}
as well as the meaning of the integration constant  $M$ are most straightforwardly 
seen within the canonical Hamiltonian treatment of (the scalar field part of) \rf{TMT}. 
Namely, upon introducing the short-hand notations:
\br
\P(B) = \pa_\m \cB^\m = \cBdot + \pa_i \cB^i \; ,
\nonu \\
\cB \equiv\cB^0 = \frac{1}{3!} \vareps^{mkl} B_{mkl} \;\; ,\;\;
\cB^i \equiv -\h \vareps^{ikl} B_{0kl} \; ,
\lab{B-short}
\er
we have for the canonically conjugated momenta $\pi_{\cB},\pi_{\cB^i}$ and $p_\vp$
w.r.t. $\cB,\cB^i$ and $\vp$:
\br
\pi_{\cB^i} = 0 \quad, \quad \pi_{\cB} = L(\vp,X) \; ,
\nonu \\
p_\vp = \bigl(\cBdot + \pa_i \cB^i + \sqrt{-g}\bigr) \partder{L}{\vpdot} \; .
\lab{can-momenta}
\er
The first relations in \rf{can-momenta} represent primary Dirac first-class
constraints and, therefore, their canonically conjugate coordinates $\cB^i$
(``electric'' components of the auxiliary ``measure'' gauge field
$B_{\m\n\l}$, cf. \rf{B-short}) are pure-gauge degrees of freedom -- 
in fact they are Lagrange multipliers 
for secondary Dirac first-class constraints (see Eq.\rf{M-constr} below).
From the second relation in \rf{can-momenta} we obtain the velocity 
$\vpdot = \vpdot (\vp,\pi_{\cB})$ as function of the canonical variables
(in the simplest case of \rf{k-essence-L} $L(\vp, X) = X - V(\vp)$):
\be
\vpdot = N^i \pa_i \vp + N \sqrt{h^{ij} \pa_i \vp \pa_j \vp
+2\bigl( V(\vp)+\pi_{\cB}\bigr)} \; ,
\lab{vpdot-eq}
\ee
where we have used the standard ADM parametrization for the Riemannian metric:
\be
ds^2 = - N^2 dt^2 + h_{ij} \bigl( dx^i + N^i dt\bigr)\bigl( dx^j + N^j dt\bigr)
\; .
\lab{ADM}
\ee
Finally, from the last relation in \rf{can-momenta} we obtain the velocity
$\cBdot$ as a function of the canonical variables. Thus, inserting 
\rf{vpdot-eq} and the second 
relation \rf{can-momenta} in the expression for the canonical scalar field Hamiltonian:
\be
\cH_m = p_\vp \vpdot + \pi_{\cB} \cBdot - 
\bigl(\cBdot + \pa_i \cB^i + \sqrt{-g}\bigr) L(\vp,X)
\lab{can-Ham-0}
\ee
we arrive at the result:
\br
\cH_m = N^i \bigl(\pa_i \vp p_\vp \bigr) 
\nonu \\
+ N \bigl\lb p_\vp \sqrt{h^{ij} \pa_i \vp \pa_j \vp
+2\bigl( V(\vp)+\pi_{\cB}\bigr)} - \sqrt{h}\pi_{\cB}\bigr\rb
\nonu \\
- \pa_i \cB^i \pi_{\cB} \; .
\lab{can-Ham}
\er
\textsl{i.e.}, scalar field canonical Hamiltonian being linear combination of 
first-class constraints only.


The last term in \rf{can-Ham} shows that $\cB^i$ are canonical Lagrange
multipliers for the secondary Dirac first-class constraints:
\be
\pa_i \pi_{\cB} = 0 \quad \longrightarrow \quad \pi_{\cB} = {\rm const}\equiv -2 M \; .
\lab{M-constr}
\ee
The latter implies that also $\cB$ (the ``magnetic'' component  of the auxiliary
``measure'' gauge field $B_{\m\n\l}$, cf. \rf{B-short}) is a
pure-gauge degree of freedom. Clearly, Eqs.\rf{M-constr} are  the canonical 
Hamiltonian analog of Eq.\rf{L-const} within the Lagrangian formalism. 
Therefore, the meaning of the arbitrary integration constant $2M$ is the
minus value of conserved Dirac-constrained canonical momentum conjugated to the
``pure-gauge'' magnetic component of the ``measure'' gauge field $B_{\m\n\l}$.
Moreover, the second term in \rf{can-Ham} shows that $M$ plays the role of a dynamically
generated cosmological constant.

Adding the well-known canonical Hamiltonian of the Hilbert-Einstein action
(upto a total derivative term \ct{regge-teitelboim}) the total canonical 
Hamiltonian of the gravity-scalar-field model \rf{TMT} is the following
linear combination of the first-class constraints:
\br
\cH_{\rm total} = N^i \cH_i + N \cH_0 - \pa_i \cB^i \pi_{\cB}
\lab{can-Ham-total} \\
\cH_i \equiv -2 D_j \pi_i^j + \pa_i \vp p_\vp \; ,
\lab{H-i} \\
\cH_0 \equiv \frac{1}{\sqrt{h}}\bigl(\pi_{ij}\pi^{ij} - \h (\pi_i^i)^2
- \sqrt{h} R^{(3)}(h) 
\nonu \\
+ p_\vp \sqrt{h^{ij} \pa_i \vp \pa_j \vp +2\bigl( V(\vp) + \pi_{\cB} \bigr)} 
- \pi_{\cB} \sqrt{h} \; .
\lab{H-0}
\er
Here $\pi^{ij}$ denote canonically conjugated momenta of the
spacial 3-dimensional ADM metric $g_{ij}=h_{ij}$,
$\sqrt{h} = \sqrt{\det\Vert h_{ij}\Vert}$, $D_i$ and $R^{(3)}(h)$ denote
covariant derivative and scalar curvature w.r.t. $h_{ij}$, respectively.

For more details about the canonical
Hamiltonian treatment of gravity-matter theories with non-Riemannian
volume-forms we refer to \ct{quintess,buggy}.

\section{Hidden Symmetry, Conservation Laws and ``Dust'' Fluid Interpretation}
\label{hidden}

We go back to the Lagrangian formalism and consider Eq.\rf{L-const}.
Multiplying its differential form
$\pa_\m L(\vp,X) \equiv \pa_\m \vp \partder{L}{\vp} + \pa_\m X \partder{L}{X} = 0$ 
by the factor $-\h g^{\m\n}\pa_\n \vp$ we get the following equivalent form
of the dynamical Lagrangian constraint \rf{L-const}:
\be
\partder{L}{\vp} 
- \frac{\pa_\m \sqrt{X}}{\sqrt{X}}\, g^{\m\n} \pa_\n \vp \partder{L}{X} = 0 \; .
\lab{L-const-1}
\ee
Inserting $\partder{L}{\vp}$ from \rf{L-const-1} into $\vp$-equations of
motion \rf{vp-eqs} we immediately rewrite the latter in the following
current-conservation law form (for later convenience we multiplied both
sides by the numerical factor $\sqrt{2}$):
\be
\pa_\m \Bigl\lb\bigl(\P(B)+\sqrt{-g}\bigr)\sqrt{2X} 
g^{\m\n}\pa_\n \vp \partder{L}{X}\Bigr\rb = 0
\lab{conserv-0}
\ee
or, equivalently, in a covariant form:
\be
\nabla_\m J^\m = 0 \quad ,\quad
J^\m \equiv \Bigl(1+\frac{\P(B)}{\sqrt{-g}}\Bigr)\sqrt{2X} 
g^{\m\n}\pa_\n \vp \partder{L}{X} \; .
\lab{J-conserv}
\ee

In fact we find a hidden 
(strongly nonlinear)
Noether symmetry of the original action \rf{TMT}
which produces $J^\m$ \rf{J-conserv} as a genuine Noether conserved current.
Indeed, the action \rf{TMT} is invariant (modulo total derivative) under the
following nonlinear symmetry transformations:
\br
\d_\eps \vp = \eps \sqrt{X} \quad ,\quad \d_\eps g_{\m\n} = 0 \; ,
\nonu \\
\d_\eps \cB^\m = - \eps \frac{1}{2\sqrt{X}} g^{\m\n}\pa_\n \vp 
\bigl(\P(B) + \sqrt{-g}\bigr)  \; ,
\lab{hidden-sym}
\er
where the short-hand notations \rf{B-short} are used. Under \rf{hidden-sym}
the action \rf{TMT} transforms as 
$\d_\eps S = \int d^4 x \pa_\m \bigl( L(\vp,X) \d_\eps \cB^\m \bigr)$. Then,
the standard Noether procedure yields precisely $J^\m$ \rf{J-conserv} as the
pertinent conserved current.

Let us particularly stress, that the existence of the hidden symmetry
\rf{hidden-sym} of the action \rf{TMT} {\em does not} depend on the specific
form of the scalar field Lagrangian \rf{k-essence-L}. The only requirement is 
that the kinetic term $X$ must be positive.

We can now rewrite $T_{\m\n}$ \rf{EM-tensor} and $J^\m$ \rf{J-conserv} in
the following relativistic hydrodynamical form (taking into account \rf{L-const}):
\be
T_{\m\n} = \rho_0 u_\m u_\n - 2M g_{\m\n} \quad ,\quad J^\m = \rho_0 u^\m \; ,
\lab{T-J-hydro}
\ee
where the integration constant 
$M$ appears as dynamically generated cosmological constant and:
\br
\rho_0 \equiv \Bigl(1+\frac{\P(B)}{\sqrt{-g}}\Bigr)\, 2X \partder{L}{X} \; ,
\lab{rho-u-def} \\
u_\m \equiv \frac{\pa_\m \vp}{\sqrt{2X}} \quad 
({\rm note} \; u^\m u_\m = -1\;) \; .
\nonu
\er
Comparing \rf{T-J-hydro} with the standard expression for a perfect fluid
stress-energy tensor $T_{\m\n} = \bigl( \rho + p) u_\m u_\n + p g_{\m\n}$,
we see that:
\be
p = - 2M \quad ,\quad \rho = \rho_0 + 2M \quad {\rm with} \;\;
\rho_0 \; {\rm as ~in ~\rf{rho-u-def}} \; ,
\lab{p-rho-def}
\ee
\textsl{i.e}, the fluid tension is constant and negative, whereas $\rho_0$ 
\rf{rho-u-def} and $2M$ are the rest-mass and internal fluid energy
densities, respectively (for general definitions, see \textsl{i.e.}
\ct{rezzola-zanotti}).

The energy-momentum tensor \rf{T-J-hydro} consists of two parts with the
following interpretation according to the standard $\L$-CDM model
\ct{Lambda-CDM-1,Lambda-CDM-2,Lambda-CDM-3} (using notations
$p = p_{\rm DM} + p_{\rm DE}$ and $\rho = \rho_{\rm DM} + \rho_{\rm DE}$ in
\rf{p-rho-def}):
\begin{itemize}
\item
Dark energy part given by the second cosmological constant term in $T_{\m\n}$
\rf{T-J-hydro}, which arises due to the dynamical constraint on the scalar field
Lagrangian \rf{L-const}, or equivalently, by \rf{p-rho-def} with
$p_{\rm DE} = -2M\, ,\, \rho_{\rm DE} = 2M$;
\item
Dark matter part given by the first term in \rf{T-J-hydro}, or equivalently,
by \rf{p-rho-def} with $p_{\rm DM} = 0\, ,\, \rho_{\rm DM} = \rho_0$
($\rho_0$ as in \rf{rho-u-def}), which in fact describes a {\em dust}.

Indeed, the covariant conservation laws for the energy-momentum tensor 
\rf{T-J-hydro} $\nabla^\m T_{\m\n} = 0$ and the $J$-current \rf{J-conserv} 
acquire the form:
\be
\nabla^\m \bigl(\rho_0 u_\m u_\n\bigr) = 0 \quad ,\quad
\nabla^\m \bigl(\rho_0 u_\m\bigr) = 0 \; ,
\lab{dust-hydro}
\ee
both of which implying the geodesic equation for the ``dust fluid'' 4-velocity $u_\n$:
\be
u_\m \nabla^\m u_\n = 0 \; .
\lab{geodesic-eq}
\ee
\end{itemize}


To conclude this section let us point out that the hidden symmetry
transformation of the scalar field (first Eq.\rf{hidden-sym}) can be
equivalently represented as a specific field-dependent coordinate shift of
the $\vp$-field (taking into account the definition of $X$ in \rf{k-essence-L}):
\br
\d_\eps \vp (x) = \eps \sqrt{X} = \vp \bigl(x+\eps \zeta_\vp (x)\bigr) - \vp (x)
\nonu \\
= \eps \zeta^\m_\vp (x) \pa_\m \vp (x) \quad ,\quad
\zeta^\m_\vp = - \frac{1}{\sqrt{2}} u^\m \; .
\lab{hidden-sym-1}
\er
Accordingly, the dust 4-velocity transforms under the hidden symmetry
(\rf{hidden-sym} or \rf{hidden-sym-1}) as:
\be
\d_\eps u^\m = \eps \bigl( g^{\m\n} + u^\m u^\n\bigr)
\frac{\pa_\n \sqrt{X}}{\sqrt{2X}} \; .
\lab{hidden-sym-2}
\ee

\section{Implications for Cosmology}
\label{cosmolog}

Let us now consider the modified gravity-scalar-field model \rf{TMT} with the
hidden symmetry \rf{hidden-sym} describing simultaneously dark matter and
dark energy in the context of cosmology. To this end let us take the
Friedmann-Lemaitre-Robertson-Walker (FLRW) metric 
(see \textsl{e.g.} \ct{weinberg-72}):
\be
ds^2 = - dt^2 + a^2(t) \Bigl\lb \frac{dr^2}{1-K r^2}
+ r^2 (d\th^2 + \sin^2\th d\phi^2)\Bigr\rb \; ,
\lab{FLRW}
\ee
and consider the associated Friedmann equations:
\be
\frac{\addot}{a}= - \frac{1}{12} (\rho + 3p) \quad ,\quad
H^2 + \frac{K}{a^2} = \frac{1}{6}\rho \quad ,\;\; H\equiv \frac{\adot}{a} \; ,
\lab{friedman-eqs}
\ee
describing the universe' evolution. In the present case we have for pressure
$p$ and  the full energy density $\rho$ the explicit expressions \rf{p-rho-def}.
Also now $\vp = \vp (t)$, so that $X = \h \vpdot^2$ and $u_\m = (1,0,0,0)$.

The $J^\m$-current conservation \rf{dust-hydro} now reads:
\be
\nabla^\m \bigl(\rho_0 u_\m\bigr) = 0 \;\; \to \;\; 
\frac{d}{dt}\bigl(a^3 \rho_0\bigr) = 0  \;\; \to \;\;
\rho_0 = \frac{c_0}{a^3} \; ,
\lab{dust-sol}
\ee
where the last relation is the typical cosmological dust solution
(see \textsl{e.g.} \ct{Lambda-CDM-2}) with $c_0 = {\rm const}$. Inserting in
\rf{dust-sol} the explicit expression \rf{rho-u-def} for $\rho_0$ we
obtain a solution for the non-Riemannian integration measure density 
$\P(B) = c_0 \bigl( 2X \partder{L}{X}\Bigr)^{-1} - a^3$, or
in the simplest case for the scalar Lagrangian ($L = \h \vpdot^2 - V(\vp)$):
\be
\P(B) = \frac{c_0}{\vpdot^2} - a^3 \; .
\lab{dust-sol-0}
\ee

Let us particularly stress, that the solution \rf{dust-sol} for the dust
(dark matter) energy density $\rho_0$ (last relation in \rf{p-rho-def})
{\em does not} depend on the specific form of the scalar Lagrangian
(cf. \rf{k-essence-L}) and the details of the dynamics of $\vp (t)$:
\be
L(\vp,X)= \frac{A_1 (\vp)}{2} \vpdot^2 + \frac{A_2 (\vp)}{4} \vpdot^4 + \cdots
- V(\vp) \; .
\lab{k-essence-L-0}
\ee

Taking into account \rf{p-rho-def} and \rf{dust-sol}, the Friedmann equations
\rf{friedman-eqs} acquire the form:
\be
\frac{\addot}{a}= - \frac{1}{12} \Bigl(\frac{c_0}{a^3} - 4M \Bigr) \quad ,\quad
\frac{\adot^2}{a^2} + \frac{k}{a^2} = 
\frac{1}{6}\Bigl(\frac{c_0}{a^3} + 2M \Bigr) \; .
\lab{friedman-eqs-0}
\ee
and, thus, the solution for $a=a(t)$ {\em does not} depend either on the specific form 
of the scalar Lagrangian \rf{k-essence-L-0} and the details of the dynamics of
$\vp (t)$. 
Exact solution for $a(t)$ of the second Eq.\rf{friedman-eqs-0} when $k=0$
was given in Ref.\ct{Lambda-CDM-1}. In the general case including radiation,
exact solutions for $a(t)$ in terms of elliptic functions can be found in
Ref.\ct{kharbediya}.

In fact, concerning the cosmological solutions of \rf{dust-sol} and
\rf{friedman-eqs-0}, the only requirement for
$L(\vp,X)$ \rf{k-essence-L-0} comes from the dynamical constraint Eq.\rf{L-const} 
on \rf{k-essence-L-0}:
\be
L(\vp,X) = - 2M \;\; \to \;\; V(\vp) > 2M \; . 
\lab{L-const-0}
\ee
In general the inequality $V(\vp) > 2M$ might define classical forbidden
regions for $\vp (t)$ (where $V(\vp) < 2M$), including turning points
$\vp_0$ (where $V(\vp_0)=2M$). In view of later applications (see discussion
about obtaining growing dark energy below) we will demand:
\be
V(\vp) > 2M \quad {\rm for ~all} \; \vp \; ,
\lab{V-inequality}
\ee
so that we will have a purely monotonic behaviour for $\vp (t)$ (cf.
\rf{vp-eq-2-0} below).

The dynamics of the scalar field $\vp (t)$ itself is given by the first-order
differential equation \rf{L-const-0}. 
Although it does not affect the cosmological solutions,
nevertheless, it is worth mentioning the following property. Taking time
derivative on both sides of \rf{L-const-0} we obtain second order evolution
equation for $\vp (t)$:
\br
\vpddot \Bigl( A_1(\vp) + \h A_2 (\vp) \vpdot^2 + \cdots\bigr) 
\nonu \\
+ \h A_1^{\pr}(\vp) \vpdot^2 + \frac{1}{4} A_2^{\pr}(\vp) \vpdot^4 + \cdots
- \partder{V}{\vp} = 0 \; .
\lab{vp-eq-2}
\er
In particular, for the standard scalar Lagrangian $L= \h \vpdot^2 - V(\vp)$
Eqs.\rf{L-const-0} and \rf{vp-eq-2} read, accordingly:
\br
\vpdot^2 = 2\bigl( V(\vp) - 2M\bigr) \;\; \to 
\nonu \\
\int_{\vp (0)}^{\vp (t)} \frac{d\vp}{\sqrt{2\bigl( V(\vp) - 2M\bigr)}} = \pm t \; , 
\lab{vp-eq-2-0} \\
\vpddot - \partder{V}{\vp} = 0 \; ,
\lab{vp-eq-2-1}
\er
where we specifically stress on the {\em opposite} sign in the force term in
the  second order $\vp$-equation of motion \rf{vp-eq-2-1}. 
Due to the dynamical constraint on
$V(\vp)$ in \rf{V-inequality} and choosing the $+$ sign the integral in \rf{vp-eq-2-0} 
yields $\vp (t)$ monotonically growing with $t$.

Let us now consider a perturbation of the initial modified-measure
gravity-scalar-field action \rf{TMT} by some additional scalar potential
$U(\vp)$ independent of the initial potential $V(\vp)$:
\br
{\wti S} = \int d^4 x \sqrt{-g}\, R + \int d^4 x \bigl(\sqrt{-g}+\P(B)\bigr) L(\vp,X)
\nonu \\
- \int d^4 x \sqrt{-g}\, U(\vp) \; .
\lab{TMT-perturb}
\er
An important property of the perturbed action \rf{TMT-perturb} is that
once again the scalar field $\vp$-dynamics is given by the unperturbed dynamical 
constraint Eq.\rf{L-const}, in particular, by Eq.\rf{vp-eq-2} or \rf{vp-eq-2-0} 
in the case of FLRW metric \rf{FLRW}. Let us strongly emphasize that the latter are
completely independent of the perturbing scalar potential $U(\vp)$.

The associated scalar field energy-momentum tensor now reads (cf.
Eqs.\rf{T-J-hydro} and \rf{p-rho-def}):
\br
{\wti T}_{\m\n} = \rho_0 u_\m u_\n + g_{\m\n}\bigl(-2M-U(\vp)\bigr)
\nonu \\
\equiv \bigl({\wti \rho} + {\wti p} \bigr) u_\m u_\n + {\wti p}\, g_{\m\n} \; ,
\lab{EM-tensor-perturb} \\
{\wti \rho} = \rho_0 + 2M + U \quad ,\quad {\wti p} = -2M - U \; ,
\lab{p-rho-def-perturb}
\er
where again notations \rf{rho-u-def} are used.

The perturbed energy-momentum \rf{EM-tensor-perturb} conservation 
$\nabla^\m {\wti T}_{\m\n} = 0$ now implies (cf. Eqs.\rf{dust-hydro} and 
\rf{geodesic-eq}):
\be
\nabla^\m \bigl(\rho_0 u_\m\bigr) - \sqrt{2X}\, \partder{U}{\vp} = 0 \quad,\quad
u_\m \nabla^\m u_\n = 0 \; .
\lab{hydro-perturb}
\ee
While we again obtain the geodesic equation for the dark matter ``fluid''
4-velocity, in the perturbed case the action \rf{TMT-perturb} does not any more possess
the hidden symmetry \rf{hidden-sym} and, therefore, the conservation of the Noether
current $J^\m = \rho_0 u^\m$ \rf{T-J-hydro} is now replaced by the first 
Eq.\rf{hydro-perturb}. In the case of FLRW metric \rf{FLRW} the latter
acquires the known form:
\be
\frac{d}{dt}\bigl(a^3 {\wti \rho}\bigr) + {\wti p}\frac{d}{dt}a^3 = 0 \; ,
\lab{hydro-perturb-0}
\ee
where the notations \rf{p-rho-def-perturb} for the total perturbed energy
density and pressure are used.

As already stressed above, the dynamics of the scalar field does not depend
at all on the presence of the perturbing scalar potential $U(\vp)$.
Therefore, if we choose the perturbation $U(\vp)$ in \rf{TMT-perturb} 
to be a growing function at large $\vp$ 
(\textsl{e.g.}, $U(\vp) \sim e^{\a\vp}$, $\a$ small positive) then, when $\vp (t)$
evolves through \rf{vp-eq-2-0} to large positive values, it (slowly) ``climbs'' the 
perturbing potential $U(\vp)$ and according to the expression 
$2M + U(\vp)$ for the dark energy density (cf. \rf{EM-tensor-perturb}), 
the latter will (slowly) grow up! Let us emphasize that in this way we obtain growing
dark energy of the ``late'' universe without any pathologies in the 
universe' evolution like ``cosmic doomsday'' or future singularities \ct{doomsday}.

Taking another example of perturbation in \rf{TMT-perturb} of the type
$U(\vp) \sim \tanh(\a\vp)$ for large $\vp$, 
then after (slowly) growing up the dark energy density $2M + U(\vp)$ will
asymptotically (for $t \to +\infty$) approach a finite constant value.

\section{Conclusions}

Let us recapitulate the main points above:
\begin{itemize}
\item
Employing a non-Riemannian volume-form (alternative covariant integration
measure density independent of the Riemannian metric) in the modified-measure
gravity-scalar-field action \rf{TMT} produces naturally a dynamically
generated cosmological constant (identified as dark energy) in the form of an 
arbitrary integration constant in solving the equations of motion \rf{L-const} 
corresponding to the auxiliary ``measure'' gauge fields.
\item
The modified-measure gravity-scalar-field action \rf{TMT} possesses a hidden
Noether symmetry \rf{hidden-sym} acting on the scalar field and the
``measure'' gauge fields (but leaving the Riemannian metric untouched), 
whose associated Noether conserved current
\rf{J-conserv} provides a relativistic hydrodynamical interpretation of the 
energy-momentum tensor \rf{T-J-hydro} describing two decoupled matter components 
-- a ``dust'' (dark matter) and a constant negative pressure (dark energy) ones. 
\item
The above unified description of dark energy and dark matter is insensitive 
w.r.t. the specific form of the scalar field Lagrangian (which might be of
higher order ``k-essence'' type) and the details of 
the underlying dynamics of the scalar field.
\item
Upon appropriate perturbing the modified-measure gravity-scalar-field action 
\rf{TMT-perturb}, which breaks the above hidden symmetry, we find a way to 
obtain growing dark energy in the present universe's epoch without evolution 
pathologies.
\end{itemize}

Straightforward quantization (\textsl{e.g.}, via functional integral) of the
scalar field action 
in \rf{TMT}, which is required to study possible quantum 
radiative instabilities within the cosmological constant problem, 
does not allow the use the standard quantum field theoretic methods
(standard perturbative expansion, Feynman diagrams and their renormalization). 
This is due to the essential nonlinearity (square root) in the
expression for the corresponding scalar field canonical Hamiltonian
\rf{can-Ham} (even in flat spacetime $N=1\, ,\, h_{ij} = \d_{ij}$) and, especially, 
because it is {\em linear} (instead of the usual
quadratic) function of the conjugated canonical momentum $p_{\vp}$. 

Canonical Hamiltonian quantization of the full gravity-scalar-field 
action \rf{TMT} were studied in \ct{frankfurt} in the {\rm reduced} case of
FLRW cosmological metric \rf{FLRW} and purely
time-dependent scalar field $\vp$. Upon appropriate 
change of variables the corresponding quantum Wheeler-DeWitt equation was reduced
(in the case of zero FLRW spacial curvature) to the Schr{\"o}dinger equation
for inverted harmonic oscillator.

\begin{acknowledgements}
We gratefully acknowledge support of our collaboration through the 
academic exchange agreement between the Ben-Gurion University in Beer-Sheva,
Israel, and the Bulgarian Academy of Sciences. 
S.P. and E.N. have received partial support from European COST actions
MP-1210 and MP-1405, respectively, as well from Bulgarian National Science
Fund Grant DFNI-T02/6. E.G. thanks Samuel Lepe for discussions.
We are also grateful to Ali Chamseddine, the referees and the editor
for useful remarks.
\end{acknowledgements}

\end{document}